\documentclass[a4paper,11pt]{article}
\pdfoutput=1 

\usepackage{jheppub}

\usepackage{datetime}

\usepackage{amsmath,graphicx,amssymb,subfigure,bbm,comment}
\usepackage[all]{xy}

\title{Entanglement Entropy of the Klebanov-Strassler with dynamical flavors}

\author{George Georgiou$^a$,}
\author{Dimitrios Zoakos$^b$}

\affiliation{$^a$ Institute of Nuclear and Particle Physics, National Center for Scientific Research Demokritos, 15310 Athens, Greece}
\affiliation{$^b$ Centro de F\'\i sica do Porto \& Departamento de F\'\i sica e Astronomia,
Faculdade de Ci\^encias da Universidade do Porto,
Rua do Campo Alegre 687, 4169--007 Porto, Portugal}

\emailAdd{georgiou@inp.demokritos.gr}
\emailAdd{dimitrios.zoakos@fc.up.pt}

\abstract{We present a detailed study of the Entanglement Entropy for the confining Klebanov-Strassler background
coupled to a large number of dynamical flavors in the Veneziano limit.
As we vary the number of the massless flavors the behavior of the entropy strongly depends on the way we fix the integration
constant of the warp factor, that is related to the glueball scale.
In the case of massive flavors, the mass of the flavor branes introduces another scale in the background and the entropy undergoes two first order phase transitions.
The competition between the glueball and the quark scales will lead to a critical point where one of the phase transitions degenerates to a second order one.
We have calculated the critical exponents and have found that they are independent of the number of flavors and different from the mean filed theory expectations.}

\begin{document}
\maketitle
\flushbottom


\section{Introduction}

One of the most inspiring conjectures in modern theoretical physics is the AdS/CFT correspondence \cite{Maldacena:1997re}
(for a set of pedagogical introductions see \cite{Ramallo:2013bua, CasalderreySolana:2011us, Edelstein:2009iv})
which, in its original form, claims the equivalence between the maximally supersymetric field theory in 4 dimensions and
type-IIB superstring theory on $AdS_5 \times S^5$. The original AdS/CFT correspondence can be generalized  to include field theories with less supersymmetry,
as well as field theories living in a different number of spacetime dimensions.
The main interest in this kind of dualities is due to the fact that they provide a means to probe the dynamics of strongly coupled gauge theories.
This is so because the duality is between strongly coupled gauge theory, on one hand, and weakly coupled string theory, on the other and 
vice-versa\footnote{This fact is at the same time a disadvantage since it is very hard to prove the conjecture. In the case of the original AdS/CFT scenario progress
on the comparison of the spectra of the two theories has been possible due to integrability \cite{Beisert:2010jr}. 
More recently, there was also some progress on the comparison of
higher point correlators, see \cite{3-point} and references therein}.

More precisely, one of the long-standing problems has been that of describing the confinement in theories like QCD.
The aim of this work is to clarify how the presence of dynamical quarks in a theory affect the phenomenon of confinement.
To this end, we choose to study the wrapped deformed conifold background (KS model  \cite{Klebanov:2000hb}) both with massless and massive
backreacting flavors (\cite{Benini:2007gx} \& \cite{Bigazzi:2008qq}).
The KS background corresponds to the low energy theory of N regular and M fractional D3-branes on the deformed conifold $z_1 z_2-z_3 z_4=\epsilon^2$.
The dual field theory is a ${\cal N}=1$ 4-dimensional gauge theory with initial gauge group $SU(N+M)\times SU(N)$, where $N=n M$ and $n$ is an integer.
The bifundamental matter fields $A,B$ interact through the superpotential $W=\epsilon^{ij}\epsilon^{kl} A_i B_k A_j B_l$ and transform as doublets
under the $SU(2)\times SU(2)$ isometry group of the deformed conifold.
This theory develops a Seiberg duality cascade which terminates after $n-1$ steps, as soon as the gauge group has been reduced to $SU(2M)\times SU(M)$.
The important point is that the KS theory exhibits confinement due to the formation of a gluino condensate $<\lambda \lambda>\sim \Lambda_{QCD}^3$.
It is apparent that the confining scale $\Lambda_{QCD}$ is related to the deformation parameter $\epsilon$ which breaks the scale invariance of the supegravity solution.

The addition of fundamental degrees of  freedom to the theory in the Veneziano limit can be achieved through the smearing of $N_f$ flavor D7-branes along
the internal directions (see \cite{Nunez:2010sf} for a review on the smearing technique and \cite{smearing} for some more smeared solutions).
This construction introduces a second scale to the theory related to the bare mass of the quarks.
It is the competition  between these two different scales which will give the interesting phenomenon of two quantum
first order phase transitions in the connected part of the Entanglement Entropy (EE). One of the phase transition is associated to the confinement-deconfinement
phase transition and is present in the unflavored KS theory, while the other is related to the  presence of dynamical quarks.
As we decrease the mass of the quarks, approaching the glueball scale, we reach to a point where one of the phase transitions degenerates to a second order one.
Beyond this point we end up with a single first order phase transition. For the details of the construction of the supergravity solutions we refer
the reader to section \ref{background}, as well as to the original literature \cite{Klebanov:2000hb,Bigazzi:2008qq, Benini:2007gx}.

From what is said above it is inferred that the observable that we will use
as an order parameter for the confinement-deconfinement
phase transition will be that of EE. In particular, we will calculate the EE between region A
and region B in the strongly coupled KS background with $N_f$ dynamical flavors.
More precisely, we will divide one of the spatial directions
into a line segment of length L, which will call region A,
and its complement, which will call region B.
Then the EE between region A and region B can be used as a measure of the effective degrees of freedom at
the energy scale $1/L$.
In \cite{ Ryu:2006bv, Ryu:2006ef, Klebanov:2007ws} a holographic prescription for the computation of the EE was
proposed, which consists of finding a minimal surface with a boundary that coincides with
that of region A.
Actually there are two different  minimal surfaces with the same end points at the boundary.
One is with two disconnected straight lines extending from each of the boundaries to inside the bulk
while the other is a connected surface that connects the two end points of region A.
In a generic confining theory, there is a critical length $L_c$ above which the disconnected lines are favored, while below it the connected curve that is favored.
Thus in a large $N$ confining theory the behavior of the EE becomes trivial in the IR ($L>L_c$) in the sense that $\partial_L S\sim N^0 <<N^2$ while
 it is of order $N^2$ for  $L<L_c$. This change of behaviour can be thought as the confinement-deconfinement
phase transition.

The plan of the paper is as follows.
In Section \ref{def-EE} we assemble all the necessary tools for the holographic computation of the EE.
In Section \ref{background} we discuss the construction of the supergravity solutions corresponding
to the inclusion of massless \& massive dynamical flavors to the KS background.
In Section \ref{masslessEE}, we analyze the behavior of the length of the strip and of the EE as we vary the number of the massless flavors.
Depending on which scale we keep fixed we will find different qualitative dependence of the afore-mentioned observables on the number of flavors.

In Section \ref{massiveEE}, we perform a similar analysis for the massive flavors.
The EE undergoes two first order phase transitions, one associated with the confinement-deconfinement phase transition while the other with the presence of the dynamical flavors.
Below some critical value of the quark mass one of the first order phase transitions degenerates to a second order one.
Calculating the critical exponents we find that their values are independent of the number of flavors and different from the mean-field theory expectations.
Finally, in Section \ref{conclusions} we draw our conclusions.


\section{Entanglement entropy computation in a confining background}
\label{def-EE}

In this section we will provide a short review of all the necessary results from \cite{Klebanov:2007ws} (see also \cite{Kol:2014nqa}), in which the authors generalized the
Ryu-Takayanagi conjecture \cite{Ryu:2006bv, Ryu:2006ef} for the gravity duals of confining large $N_c$ gauge theories.

In a quantum field theory the EE between two complementary spatial regions A \& B is the entropy measured
by an observer in region A, who has no access to the degrees of freedom in region B. In \cite{Klebanov:2007ws, Ryu:2006bv, Ryu:2006ef} a holographic computation
was proposed, which consists of finding a minimal surface with a boundary that coincides with that of region A.
Then, the EE between region A and the complementary region B is given by the following expression
\begin{equation}\label{KKM_conjecture}
 S \, = \, \frac{1}{4 \, G_{10}} \, \int_{\Sigma}  d^8 \xi \, e^{-2\Phi} \,\sqrt{\hat{G}_{8}} \, ,
\end{equation}
where $G_{10}$ is the ten-dimensional Newton constant \& $\hat{G}_{8}$ is the induced string frame metric on $\Sigma$.
The calculation of the EE is realized through the minimization of the action \eqref{KKM_conjecture} over all the surfaces that approach the boundary of the entangling surface.

A generic expression of the string frame metric for the type of  backgrounds that we will consider is
\begin{equation}\label{background}
 ds^2 \,= \, \alpha (\tau) \, \left[  \beta(\tau) \, d\tau^2 \, + \, dx^{\mu} dx_{\mu}  \right] \, + \,
g_{ij} \, d\theta^i \, d\theta^j \, ,
\end{equation}
where $x^{\mu}$ $\left(\mu=0,1,\dots,d \right)$ parameterize $\mathbb{R}^{d+1}$,
$\theta^{i}$ $\left(  i=d+2,\dots,9 \right)$ are the $8-d$ internal directions and $\tau$ is the holographic radial coordinate\footnote{$\tau_{\Lambda}$
is zero for the KS geometry while there is a finite UV cut-off (due to the presence of a Landau pole) when we consider the KS background with massless/massive dynamical flavors.}
\begin{equation}
\tau_{\Lambda}<\tau<\tau_{*} \, .
\end{equation}
The background is supplemented by a dilaton (that we will denote by $\Phi$) and some fluxes (which will not enter in the EE calculation).
The volume of the internal manifold is
\begin{equation}
V_{int}=\int  \prod_{i=1}^{8-d} d\theta^i\sqrt{\det[g_{ij}]} \, .
\end{equation}

In \cite{Klebanov:2007ws} the authors focused their attention on entangling surfaces that are strips of length $L$.
Analyzing the equation of motion they found that it has two independent solutions corresponding to two local minima of the action \eqref{KKM_conjecture}.
The first is a disconnected surface, consisting of two cigars that extend in $\mathbb{R}^{d-1}$ and separate in the remaining direction $\mathbb{R}^{d}$ by a distance $L$.
The second is a connected surface, where the two cigars are connected with a tube whose width depends on $L$.
The length of the line segment for the connected solution is
\begin{equation}\label{length}
L (\tau_0) \, = \, 2   \int_{\tau_0}^{\tau_{*}} d\tau \, \sqrt {\frac{\beta(\tau) }{\frac{  H(\tau)  }{ H(\tau_0)  } \, - \, 1 }   } \, ,
\end{equation}
where $\tau_0$ is the minimal value of $\tau$ along the connected surface in the bulk and we have also defined the following
quantity\footnote{The complicated integral in \eqref{length} can be approximated unexpectedly well with a combination of the background functions that appears in
equation (3.1) of \cite{Kol:2014nqa}. This claim has been checked for some of the results that we will present in the rest of this paper.}
\begin{equation}\label{H}
H(\tau) \, = \, e^{-4\Phi} \, V_{int}^2 \, \alpha^d \, .
\end{equation}
The EE of the connected surface as a function of $\tau_0$ is given by
\begin{equation}\label{EEstripC}
 S_{C}(\tau_0) \, = \, \frac{V_{d-1}}{2 \, G_{10}}  \int _{\tau_0}^{\tau_{*}} d\tau \,
\sqrt{\frac{\beta(\tau)H(\tau)}{1-\frac{H(\tau_0)}{H(\tau)}}} \, ,
\end{equation}
while the EE of the disconnected surface is independent of $\tau_0$ and given by
\begin{equation}\label{EEstripD}
S_{D}(\tau_0) \, = \, \frac{V_{d-1}}{2 \, G_{10}}  \int _{\tau_{\Lambda}}^{\tau_{*}} d\tau \, \sqrt{\beta(\tau) \, H(\tau)} \, .
\end{equation}
Even if the EE is in general UV divergent, the difference between \eqref{EEstripC} and \eqref{EEstripD} is finite
\begin{equation}\label{EE_difference}
\frac{2 \, G_{10}}{V_{d-1}} \, S(\tau_0) \, \equiv \, \frac{2 \, G_{10}}{V_{d-1}} \, (S_C \, - \, S_D) \, = \,
\int _{\tau_0}^{\tau_{*}} d\tau \, \sqrt{\frac{\beta(\tau) \, H(\tau)}{1 \, - \, \frac{H(\tau_0)}{H(\tau)}}} \, - \,
\int _{\tau_{\Lambda}}^{\tau_{*}} d\tau \, \sqrt{\beta(\tau) \, H(\tau)} \, ,
\end{equation}
and it is this quantity that we will calculate in the specific case of the KS background with massless/massive dynamical flavors.


\section{Klebanov--Strassler background with dynamical flavors}
\label{background}

In this section we will review the two known solutions of  \cite{Benini:2007gx} and
\cite{Bigazzi:2008qq, Conde:2011rg} for
the addition of massless and massive flavors to the Klebanov-Strasser background  \cite{Klebanov:2000hb}.
We will arrange the constants in such a way that the limit from the backreacted to the unflavor solution
is smooth, something that will be useful when we will perform the calculation of the EE.


\subsection{Addition of massless dynamical flavors}
\label{massless-flavors}

The background we will construct accounts for the addition of fully backreacted (fractional and regular)
D3-branes and (smeared flavor) D7-branes on the deformed conifold.  The action consists of the
usual color part (type IIB supergravity) and the flavor part (Dirac-Born-Infeld \&  Wess-Zumino terms).
The ansatz for the metric (in the Einstein frame) we will adopt is the following
\cite{Benini:2007gx, Bigazzi:2008qq}
\begin{eqnarray}  \label{metric}
ds^2 &=& h^{-1/2}(\tau)\,dx_\mu\,dx^{\mu} \, + \, h^{1/2}(\tau)\,ds_6^2\quad  \text{with} \quad \\ [4pt]
ds_6^2 &=& \frac{1}{9} \,e^{2G_3} \,\left(d\tau^2 +  g_5^2 \right) +
e^{2G_1} \, \left(1 - \frac{1}{\cosh \tau}\right) \left(g_1^2 + g_2^2 \right) +
e^{2G_2} \left(1 + \frac{1}{\cosh \tau}\right) \left(g_3^2 + g_4^2 \right) \,,
\nonumber
\end{eqnarray}
where
\begin{eqnarray}
& g_1 \,= \, \frac{1}{\sqrt{2}} \, \left( \omega_2 - \sigma_2 \right) \, , \qquad
g_3 \, = \, \frac{1}{\sqrt{2}} \, \left( \omega_2 + \sigma_2 \right) \, ,&
\nonumber \\[4pt]
& g_2= \frac{1}{\sqrt{2}} \, \left( - \omega_1 + \sigma_1 \right)  \, , \qquad
g_4 \, = \, \frac{1}{\sqrt{2}} \, \left( \omega_1 + \sigma_1 \right)  \quad \& \quad
g_5 \, = \, \omega_3 + \sigma_3  \, , &
\end{eqnarray}
with
\begin{eqnarray}
&&\omega_1\,=\,\sin{\psi} \sin{\theta_2}\,
d\varphi_2\,+\,\cos{\psi}\,d\theta_2 \, , \qquad
\omega_2\,=\,-\cos{\psi}
\sin{\theta_2}\, d\varphi_2\,+\,\sin{\psi}\,d\theta_2 \, ,
\nonumber\\
&& \omega_3\,=\,d\psi\,+\,\cos{\theta_2}\,d\varphi_2 \, ,  \quad
\sigma_1\,=\,d\theta_1 \, , \quad
\sigma_2\,=\,\sin{\theta_1} \, d\varphi_1 \quad \& \quad
\sigma_3\,=\,\cos{\theta_1}\,d\varphi_1 \, .
\end{eqnarray}
The range of the angles is $\psi \in [0,4\pi)$, $\varphi_i \in [0,2\pi)$, $\theta_i \in [0,\pi]$ and $\tau\in [0,\infty)$.
The expressions for the dilaton and the (RR and NS) forms are (in units of $g_s=1$ \& $\alpha'=1$)
\begin{eqnarray}
F_5&=&d h^{-1}(\tau) \, \wedge dx^0\wedge\cdots\wedge dx^3\,+\,{\rm Hodge\,\,dual} \,,\qquad
\Phi \, = \, \Phi(\tau)\,, \nonumber \\[4pt]
B_2 &=& \frac{M}{2} \Bigl[ f\, g^1 \wedge g^2\,+\,k\, g^3 \wedge g^4 \Bigr]\,, \qquad
F_1\, = \, {N_f \over 4\pi}\,\,g^5\,,
\nonumber \\[4pt]
H_3&=&  \frac{M}{2} \, \Bigl[ d\tau \wedge (\dot f \,g^1 \wedge g^2\,+\,
\dot k\,g^3 \wedge g^4)\,+\,{1 \over 2}(k-f)\, g^5 \wedge (g^1\wedge g^3\,+\,g^2 \wedge g^4) \Bigr]\,,
\nonumber \\[4pt]
F_3&=& \frac{M}{2} \Big\{ g^5\wedge
\Big[ \big( F+\frac{N_f}{4\pi}f\big)g^1\wedge g^2 +
\big(1- F+\frac{N_f}{4\pi}k\big)g^3\wedge g^4 \Big] +
\dot F d\tau \wedge \big(g^1\wedge g^3 + g^2\wedge g^4   \big)\Big\} \,,
\nonumber
\end{eqnarray}
where $M$ is the fractional D3-brane Page charge.
The $G_i=G_i(\tau)$ ($i=1,2,3$), $h=h(\tau)$, $f=f(\tau)$, $k=k(\tau)$ and $F=F(\tau)$ are
unknown functions of the radial coordinate, while the dot denotes derivative with respect to $\tau$.

The Bianchi identities for $H_3$ and $F_3$ are automatically satisfied by the proposed ansatz while the one
for $F_5$ can be reduced to the following first order differential equation
\begin{equation} \label{BPS_h}
4 \, \dot h\, e^{2G_1+2G_2}\,=\,- \,  M^2\Big[ f \, - \, \left(f \, - \, k \right) F \, + \, {N_f \over 4\pi} \, f \, k \Big]  \, .
\end{equation}
Applying the fermionic supersymmetric variations we obtain a system of  BPS equations that fully
determine the unknown functions of the background. The BPS equations for the functions of the
metric are
\begin{eqnarray} \label{BPS_G}
\dot G_1&=&\frac{1}{18} \, e^{2G_3-G_1-G_2} \, + \, \frac{1}{2} \, e^{G_2-G_1} \, - \,
\frac{1}{2} \, e^{G_1-G_2}\,,
\nonumber \\
\dot G_2&=&\frac{1}{18} \, e^{2G_3-G_1-G_2} \, - \, \frac{1}{2} \, e^{G_2-G_1} \, + \,
\frac{1}{2} \, e^{G_1-G_2}\,,
\nonumber \\
\dot G_3 &=& - \, \frac{1}{9} \, e^{2G_3-G_1-G_2} \, + \, e^{G_2-G_1} \, - \,
\frac{N_f}{8\pi} \, e^{\Phi}\,,
\end{eqnarray}
and for the dilaton it is
\begin{equation} \label{Dil_sol}
{\dot \Phi} \, = \, \frac{N_f}{4\pi} \, e^{\Phi} \quad \Rightarrow \quad
e^{-\Phi} \, = \, 1 \, + \, \frac{N_f}{4\pi} \, \left( \tau_* \, - \, \tau \right)\, ,
\end{equation}
where we have introduced an extra scale $\tau_*$, at which we fix the value of the dilaton.
Notice that the solution in \eqref{Dil_sol} is defined up to $\tau < \tau_{LP}$, where
$\tau_{LP} \, = \,  \tau_* \, + \, \left(\frac{N_f}{4\pi}\right)^{-1}$  is the point that the dilaton blows up.
In order to write in a compact form the solution of the BPS system in \eqref{BPS_G},
we introduce the following set of auxiliary functions $\Lambda(\tau)$ \& $\eta(\tau)$\footnote{We did not integrate the differential equation for
$\eta$ in \eqref{def_lambda} in order to make smooth contact with the massive case that we will present in the
next subsection. If we integrate and substitute the expression for $\eta$ in the one for $\Lambda$
we have
\begin{equation} \label{def_lambda_massless}
\Lambda(\tau)\,\equiv\, \frac{1}{2^{1/3} \, \sinh \tau}
\Bigg[\sinh 2\tau -  2 \tau + \frac{N_f}{8 \pi} \, \frac{\cosh 2 \tau - 2 \tau^2}{1 + \frac{N_f}{4\pi} \, \left( \tau_*  - \tau \right)}  \Bigg]^{{1\over 3}} \, .
\end{equation}}
\begin{equation} \label{def_lambda}
\Lambda(\tau)\,\equiv\, \frac{1}{2^{1/3} \, \sinh \tau}
\Big[ \sinh 2\tau - 2 \tau + \, \eta(\tau) \Big]^{{1\over 3}}
\quad \text{with} \quad
\frac{d}{d \tau} \left( e^{-\Phi} \eta \right) \, = \,  \frac{N_f}{4 \pi} \,\left(\sinh 2 \tau - 2 \tau \right) \, .
\end{equation}
The metric functions $G_i$ in terms of the expression for the dilaton \eqref{Dil_sol} are
\begin{equation} \label{G-sol}
e^{2G_1}\,=\,{\epsilon^{{4\over 3}} \over  4} \, e^{-{\Phi \over 3}} \,
{\sinh^2\tau\over \cosh\tau}\,\Lambda(\tau)\,,
\quad
e^{2G_2}\,=\,{\epsilon^{{4\over 3}}  \over 4}\, e^{-{\Phi \over 3}} \,
\cosh\tau\,\Lambda(\tau)\, ,
\quad
e^{2G_3}\,=\,{3 \over 2} \,\epsilon^{{4\over 3}}\,
{e^{-{\Phi \over 3}} \over \Lambda(\tau)^2}\,\,,
\end{equation}
where $\epsilon$ is an integration constant.
Supersymmetry imposes the following constraints on the functions that appear in the
expressions for the fluxes
\begin{equation} \label{BPS_kfF}
\dot{k}\,=\,e^{\Phi}\,\Big(F + {N_f\over 4\pi}\,f \Big)\, \coth^2{{\tau\over 2}}\, , \quad
\dot{f}\,=\,e^{\Phi}\,\Big(1- F + {N_f\over 4\pi}\,k \Big)\, \tanh^2{\tau\over 2}\, , \quad
\dot{F}\,=\,{1 \over 2} \, e^{-\Phi} (k-f) \,,
\end{equation}
which are solved as follows
\begin{equation} \label{kfF_sol}
f = e^{\Phi}\,{{\tau \coth{\tau}-1} \over {2 \sinh{\tau}}}(\cosh{\tau}-1)\, ,
\quad
k=e^{\Phi}\,{{\tau \coth{\tau}-1} \over {2 \sinh{\tau}}} (\cosh{\tau}+1)\,,
\quad
F={{\sinh{\tau}-\tau} \over {2 \sinh{\tau}}}\, ,
\end{equation}
where the dependence on the number of flavors is through the equation for the dilaton \eqref{Dil_sol}.
Combining \eqref{G-sol} with \eqref{kfF_sol} and \eqref{BPS_h} we can immediately solve for the
warp factor $h(\tau)$.  The integration constant can be fixed in different ways and in section
\ref{masslessEE} we will see the effect of two alternative possibilities on the qualitative behavior of the
EE.


\subsection{Addition of massive dynamical flavors}
\label{massive-flavors}

The BPS equations between the massless \cite{Benini:2007gx} and the massive \cite{Bigazzi:2008qq} case are unchanged, up to the substitution
\begin{equation} \label{massiveNf}
N_f \quad \Rightarrow \quad N_f(\tau) \, .
\end{equation}
This is the case since the supersymmetric variations contain only expressions of the $F_{i}$`s and not of their derivatives,
despite the difference between the modified Bianchi identities of the massive and the massless case. The extra ingredient in order
to solve the BPS equations is to determine the distribution function of the D7-brane charge density by a microscopic calculation, which was performed in
\cite{Bigazzi:2008qq, Conde:2011rg}. The outcome of that analysis is a very complicated expression that looks like a smoothed-out Heaviside step function
\begin{equation} \label{step}
N_f(\tau) \, = \, N_f \, \  \Theta (\tau \, - \tau_q) \, ,
\end{equation}
where $ \tau_q$ is related to the dynamical quark mass\footnote{At the probe level the radius of the spherical cavity, that the flavor branes distribute, correspond to the bare mass.
When backreaction is taken into account see  \cite{Filev:2011mt,Erdmenger:2011bw,Magana:2012kh}.}. Comparison between calculations using the exact expression of the brane distribution function and the
Heaviside approximation \eqref{step}, of the quark antiquark potential \cite{Bigazzi:2008zt}  and of the meson spectrum \cite{Bigazzi:2009gu} in the case of the
massive Klebanov-Strassler and Klebanov-Witten backgrounds, have shown that qualitatively the physics remains unchanged.
Based on this observation we will perform the calculations of the EE using the heaviside approximation, since additionally using the exact expression of the
brane distribution function makes the computation extremely time expensive.

The smeared set of massive flavor branes produces a spherical cavity that effectively splits the solution
into two regions. In the inner region, where $\tau \in [0, \tau_q]$, the effective charge for the D7-branes is zero
and imposing regularity at the level of the cavity the solution is a mild modification of the known
KS unflavored result. In the outer region, where $\tau >  \tau_q$, we have to solve the new system of BPS
equations with the substitution \eqref{massiveNf}.

Starting with the dilaton equation \eqref{Dil_sol} we can easily integrate and, after imposing regularity at the cavity,
obtain the following expression
\begin{equation}\label{Dil-sol-massive}
  e^{- \, \Phi} =
  \begin{cases}
  1 \, + \, \frac{N_f}{4\pi} \, \left( \tau_* \, - \, \tau_q \right)         & \text{if } \tau \leq \tau_{q} \\
  1 \, + \, \frac{N_f}{4\pi} \, \left( \tau_* \, - \, \tau \right)     &  \text{if } \tau > \tau_{q} \, .
  \end{cases}
\end{equation}
Following the same line as in the massless case we are able to solve the massive BPS system and the only function that needs to be
determined is $\eta(\tau)$. From \eqref{def_lambda} we can see that this function has a constant value inside the cavity, where the unflavored solution lives.
Imposing continuity at $\tau=\tau_q$ we can fix this constant to zero, so the expression for  $\eta(\tau)$ becomes (in the Heaviside approximation)
\begin{equation}\label{eta-sol-massive}
  \eta(\tau) =
  \begin{cases}
  0         & \text{if } \tau \leq \tau_{q} \\
  \frac{N_f}{8 \pi} \, \frac{\cosh 2 \tau - 2 \tau^2 -\cosh 2 \tau_q + 2 \tau_q^2}{1 + \frac{N_f}{4\pi} \,
  \left( \tau_*  - \tau \right)}  &  \text{if } \tau > \tau_{q} \, .
  \end{cases}
  \end{equation}
Taking into account that the expressions for the fluxes are still given by  \eqref{kfF_sol},
having determined the dilaton and the function $\eta$, from \eqref{Dil-sol-massive} \& \eqref{eta-sol-massive}
respectively, completely solves the problem. As in the massless case the expression for the warp factor
$h(\tau)$ is obtained by integrating \eqref{BPS_h} and imposing continuity at the level of the cavity.

In order to check the regime of validity of the supergravity dual of the current section, we can study, in the string frame metric,
the Ricci scalar $R$ and the square of the Ricci tensor $R_{MN}R^{MN}$. It is not difficult to check that
\begin{equation}
e^{\Phi} \sim {1 \over N_f} \quad \&  \quad  h \sim {M^2 \over N_f} \quad \Rightarrow \quad R \sim {1 \over \sqrt{h \, e^{\Phi}}} \sim {N_f \over M} \, ,
\end{equation}
and therefore the solution is reliable in the interval $1\ll N_f\ll M$. This calculation is valid for both the massless \& the massive solution.


\section{Entanglement entropy \& massless flavors}
\label{masslessEE}

In this section we will place a strip of length $L$ at a finite UV cut off scale of the Klebanov-Strassler geometry with massless dynamical flavors
and calculate the EE difference, as it is defined in \eqref{EE_difference}. Since we want to emphasize the effect of the dynamical flavors we fix a comparison scheme
in which we vary the number of flavors, while we will keep fixed the UV scale $\tau_*$ and the integration constant $\epsilon$ (defined in \eqref{Dil_sol} \& \eqref{G-sol} respectively).
An alternative choice would be to keep fixed the Landau pole, as it is defined below \eqref{Dil_sol}, but to ensure a smooth zero flavor limit (unflavored solution)
 we need a flavor dependent Landau pole\footnote{For the possible physical implications of the different comparisons schemes see \cite{Filev:2014nza}.}.

\begin{figure}[h] 
   \centering
   \includegraphics[width=7.5cm]{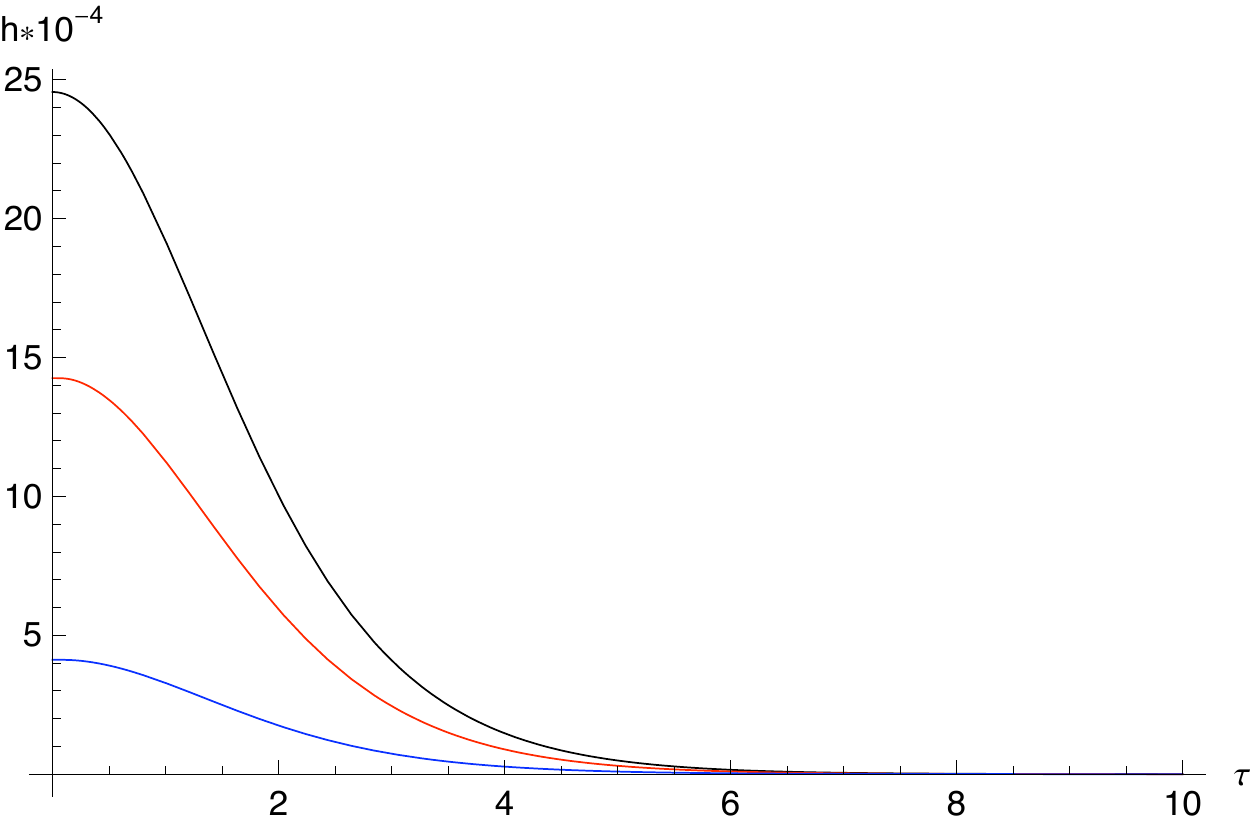}
   \includegraphics[width=7.5cm]{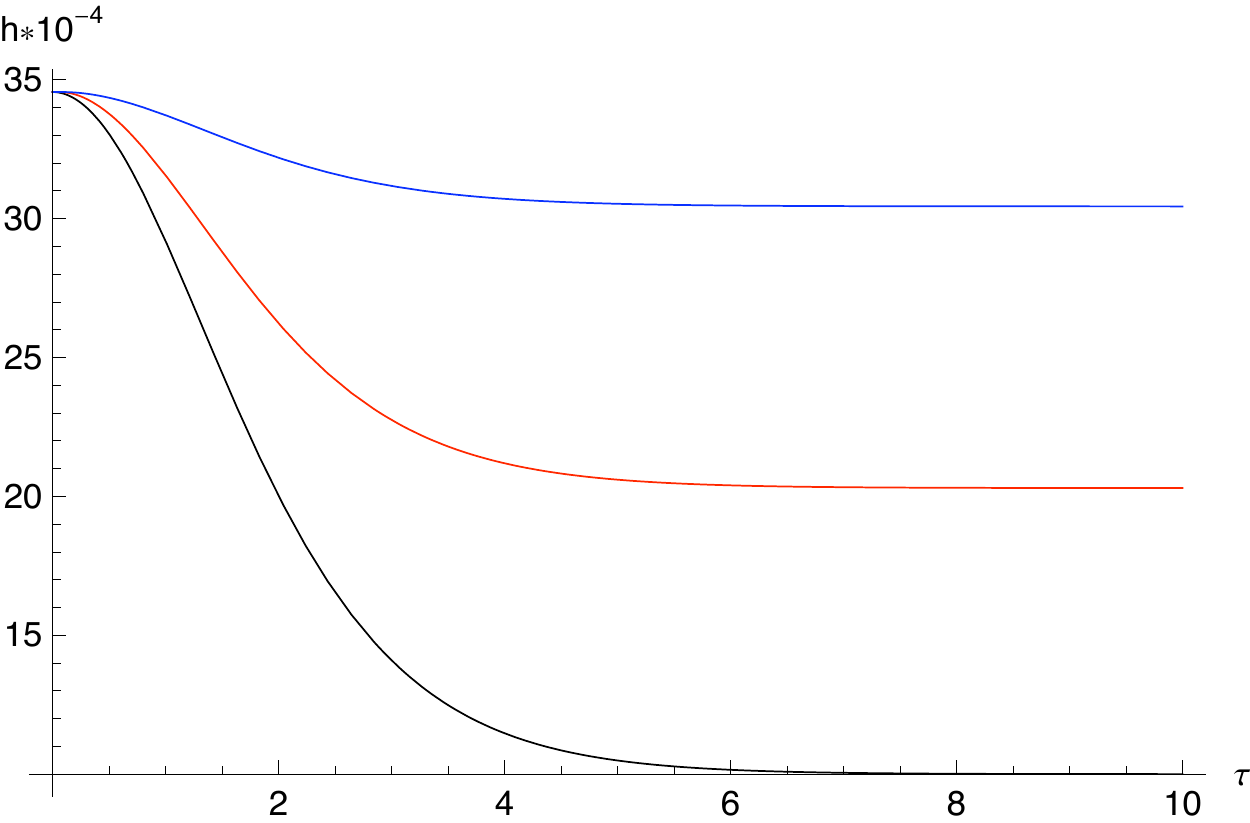}
   \caption{We have plotted the warp factor as a function of the holographic coordinate for the two different
   choices of fixing the integration constant $h_0$. The black curve is for zero flavors and in both cases
   matches the computation of \cite{Klebanov:2007ws}.
   The red(blue) curve is for $N_f=0.1$($N_f=0.5$). In the numerical evaluation we have used $\epsilon=1$, $M = 10^{8/3}$  \& $\tau_*=50$.}
   \label{fig:1}
\end{figure}

As we mentioned in subsection \ref{massless-flavors} another constant that needs to be fixed
is the integration constant that comes from the equation of the warp factor $h(\tau)$, namely $h_0$.
This constant determines the scale for the glueball and KK masses, since $m_{glue} \sim 1/\sqrt{h_0}$.
We will consider two specific choices of fixing the constant $h_0$\footnote{In \cite{Bigazzi:2008qq}
the authors examined the effect of two alternative possibilities on the Wilson loops calculation.}:
\begin{itemize}
\item {\bf Case I.} We fix $h_0$, in analogy with the unflavored
KS solution \cite{Klebanov:2000hb}, by requiring that $h(\tau)$ goes to zero at some finite UV cut off
(before reaching the Landau pole that the supergravity approximation is no longer valid) for any number
of flavors\footnote{See more about the effect of such a choice for the UV cutoff on the IR physical quantities in \cite{Bigazzi:2009gu}.
EE is even more less affected, since in \eqref{EE_difference} we are subtracting the contribution from the UV}. In this case $h_0(N_f)$, as can be seen from the left plot of figure \ref{fig:1}, is a monotonically decreasing function and as a consequence the glueball mass increases with the number of flavors.
\item {\bf Case II.} We fix $h_0$ by imposing that the value of $h(\tau)$ at the origin is the same for any value of $N_f$, see the right plot of figure \ref{fig:1}.
Here we invert the strategy that we followed in Case I, and since the numerical value of the glueball mass may depend on other parameters of the geometry besides $h_0$, we will
obtain a glueball mass that decreases with the number of flavors. The price we have to pay for such a choice is that
the warp factor is finite in the UV, which amounts to deforming the QFT dual with the insertion of an irrelevant operator (see the discussion in \cite{Elander:2011mh}).
\end{itemize}
In order to numerically perform the computation, we need to specify the exact expressions of the functions $H(\tau)$ \& $\beta(\tau)$ that appear in the expressions of
section \ref{def-EE}\footnote{Note that the dilaton has disappeared from $H(\tau)$ since we are working in the string frame.}
\begin{equation}\label{H-beta-sol}
H(\tau) \, = \,\frac{64 \pi^6}{9} \, h(\tau) \, e^{4 G_1 + 4 G_2 + 2 G_3}
\quad \& \quad
\beta(\tau) \, = \,\frac{1}{9} \, h(\tau) \, e^{2 G_3}  \, .
\end{equation}
\begin{figure}[h] 
   \centering
   \includegraphics[width=7.5cm]{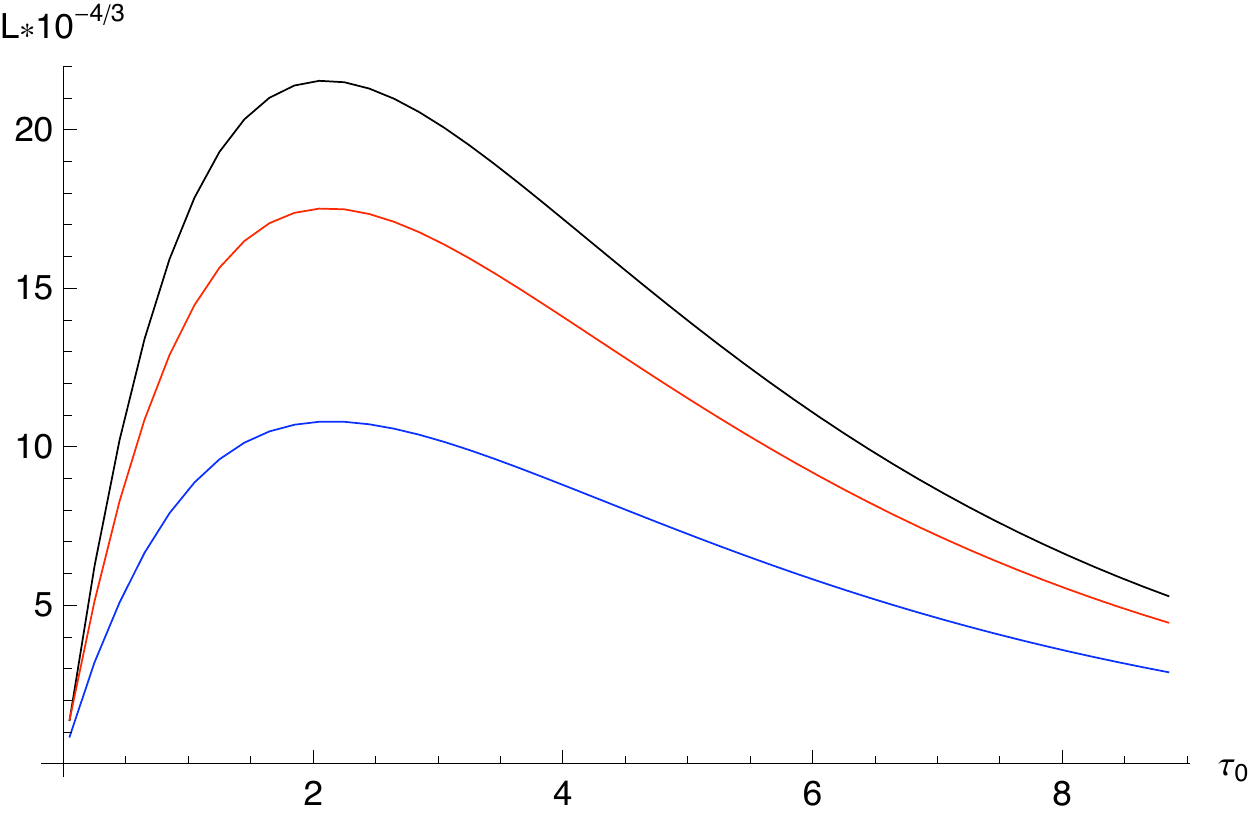}
   \includegraphics[width=7.5cm]{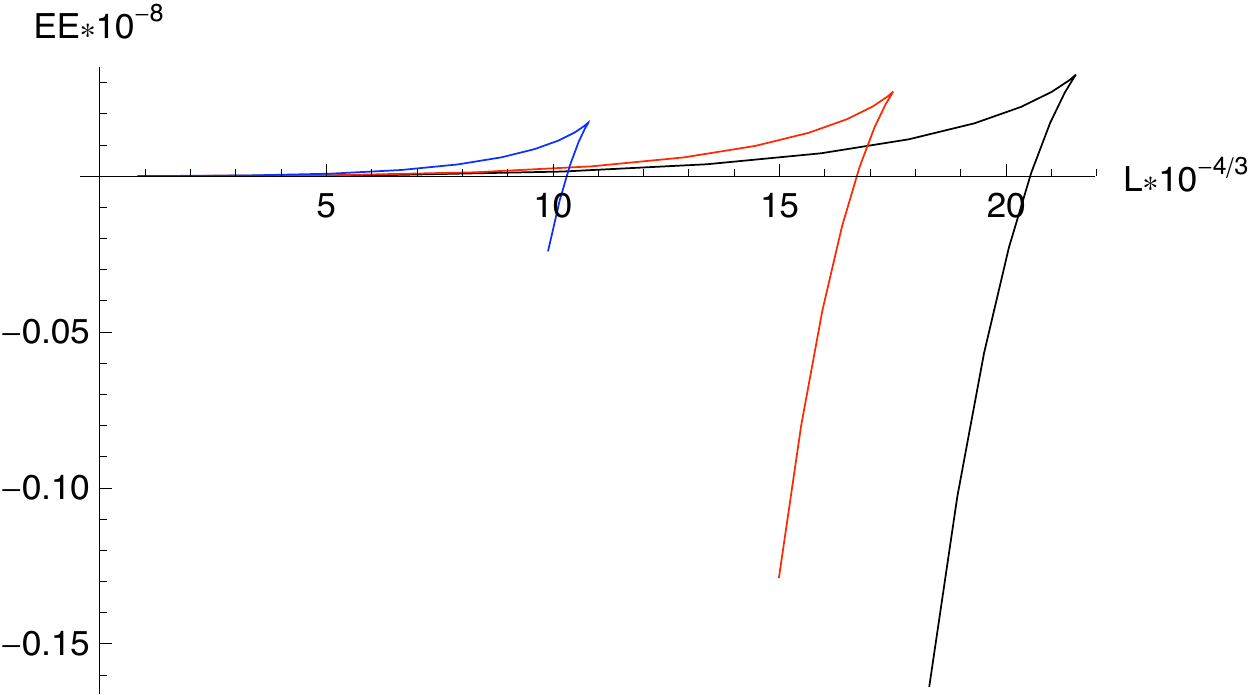}
   \caption{On the left part of the figure it is the plot of the length of the strip as a function of $\tau_0$, which is the minimal value of $\tau$ in the bulk of the geometry.
   In the right part of the figure it is the plot of the EE as a function of $L$. The black curve is for zero flavors and reproduces the computation of \cite{Klebanov:2007ws}.
   The red(blue) curve is for $N_f=0.1$($N_f=0.5$).}
   \label{fig:2}
\end{figure}
As can be seen from both figures \ref{fig:2} \&  \ref{fig:3} the EE can either be positive or negative depending on the length of the strip.
When the EE is positive the solution is a disconnected surface and when it is negative a connected. The transition between these two phases is characteristic in confining theories
and first appeared in the literature in \cite{Klebanov:2007ws} (see also \cite{Kol:2014nqa}).

The connected solution exists only until a maximum value for the length of the strip  (the point where $L'(\tau)=0$ \&  $L''(\tau)<0$).
As can be seen from the plots of both figures \ref{fig:2} \&  \ref{fig:3}, there are two possible branches for the connected solution.
The upper branch is an unstable solution while the lower is a stable one.
The double valueness corresponds to a first order phase transition between the connected and the disconnected solution,
at the critical point $L_{crit}$ (it is the point at which the EE is zero).
In \cite{Klebanov:2007ws} the authors claimed that every change in the trend of the function $L (\tau_0)$
is a sign for a confinement/deconfinement phase transition.
In other words, every peak of the function $L(\tau_0)$ corresponds to a phase transition. 
We will support this claim with our computations both in the massless \& massive KS background.

In figure \ref{fig:2} we plot the  length of the strip and the EE as a function of $\tau_0$ as well as the EE as a function of $L$.
As can be seen from the left plot of this figure, the critical length $L_{crit}$ decreases as the number of flavors
increases. Naively, from a weak field theory point of view, that is what we should expect since increasing the number of flavors will in turn increase screening.
From another point of view increasing the number of flavors, increases the beta function and decreases the QCD scale $\Lambda_{QCD}$.
Since we expect the location of the transition (between the connected to the disconnected configuration) to happen at $L_{crit}\, \sim \Lambda^{-1}_{QCD}$,
the critical length should increase with the number of flavors.
This was observed  in \cite{Kim:2013ysa}  for the behavior of the critical length in the computation of the EE that was performed in the
MN background with massless dynamical flavors \cite{HoyosBadajoz:2008fw}.
The reason behind introducing an alternative way of fixing the constant $h_0$ (Case II) was to show that it is possible even in the massless KS background to reproduce such a
behavior for the critical length.

Before analyzing the behavior of the EE when we fix $h_0$ along the lines of Case II, we would like to point out a couple of things about the calculation of the EE in the
(massless) MN vs. KS backgrounds and the importance of the value of $h_0$ for the appearance of a phase transition.
Comparing the calculations for the EE between the KS and the MN backgrounds (without flavors), performed in \cite{Klebanov:2007ws} and \cite{Kim:2013ysa} respectively,
it is clear that while in the KS case there is a phase transition in the MN case there is not, since $L$ is a monotonically increasing function.
In \cite{Kim:2013ysa} in order to reproduce confining behavior they were forced to introduce,
by hand, a UV cut-off scale. On the other hand performing the computation of the EE in the KS background it is clear that the choice of fixing $h_0$ is very important.
Had they chosen to fix it in a way that the warp factor is constant at infinity, the phase transition in the EE immediately disappears.

Combining these two observations we conclude that if we want to fix $h_0$ along Case II and still be able to reproduce confinement,
we should exploit the presence of the flavor driven UV cut-off scale. In this way confinement appears close to the UV cut-off scale, in a way that seems artificial.
There are ways to circumvent this by performing a UV completion of the background, along the lines that are extensively discussed in \cite{Kol:2014nqa}.

\begin{figure}[h] 
   \centering
   \includegraphics[width=7.5cm]{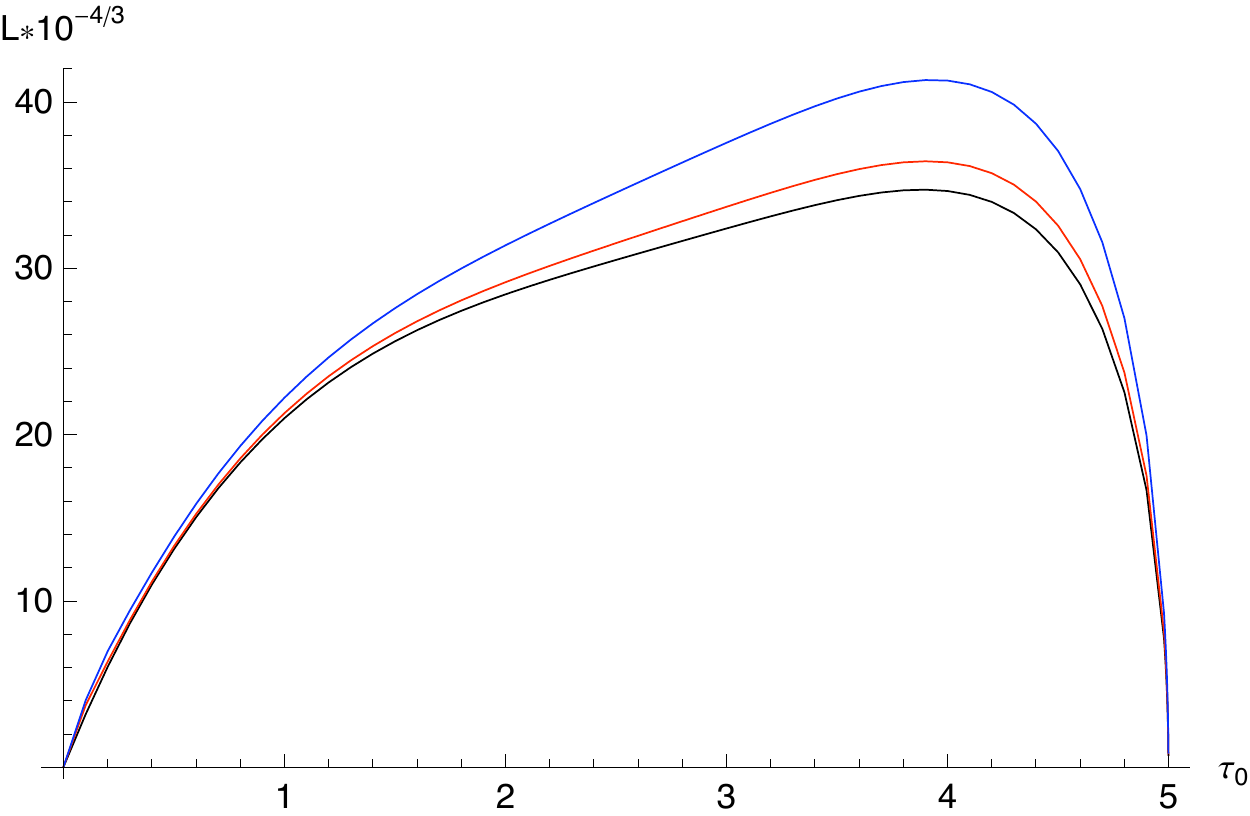}
   \includegraphics[width=7.5cm]{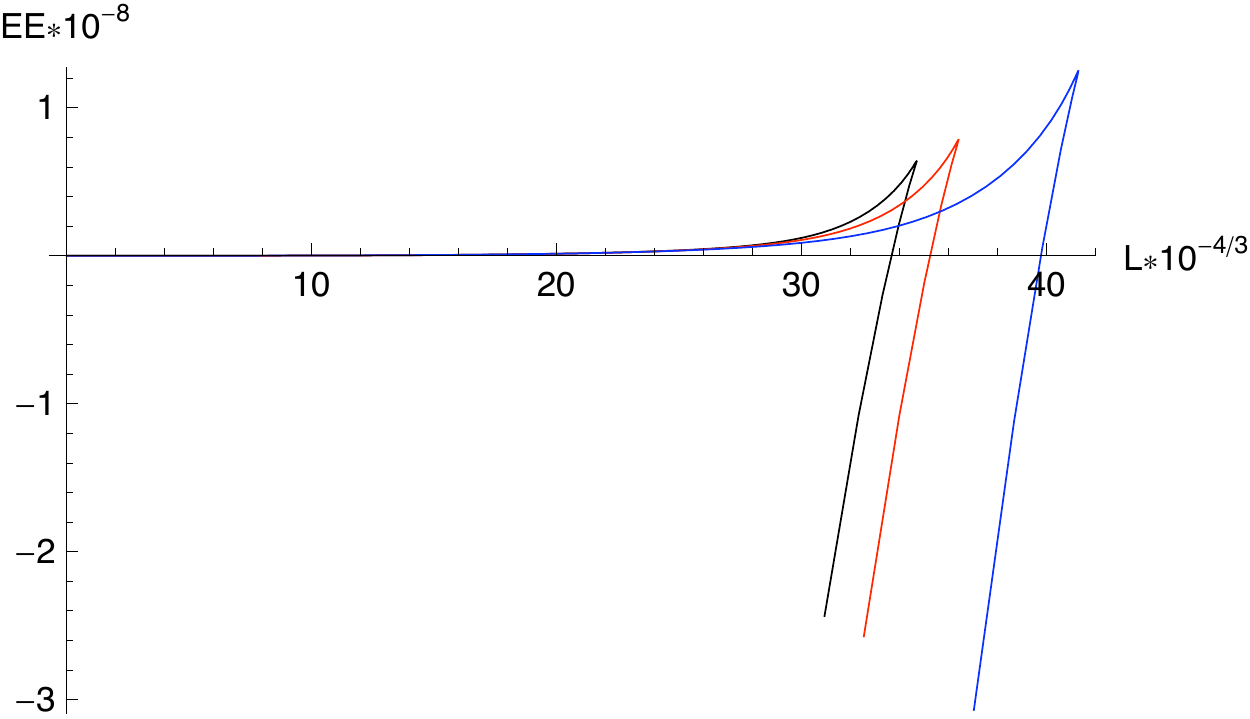}
   \caption{This is the analogue of figure \ref{fig:2}, when we fix $h_0$ as it is described in Case II.
   The red(blue) curve is for $N_f=0.1$($N_f=0.5$) }
   \label{fig:3}
\end{figure}

In figure \ref{fig:3} we plot the  length of the strip and the EE as a function of $\tau_0$ as well as the EE as a function of $L$, but now $h_0$ is fixed as it is described in Case II.
In this way, contrary to what it is described in figure  \ref{fig:2}, the critical length increases and the glueball mass decreases with the number of flavors.
The glueball scale is set by hand through the choice of $\tau_*$.

In the following paragraph we will study the EE in the massive KS background and focus the analysis on quantum phase transitions.
For this reason we will not consider Case II and all the calculations will be performed in Case I.


\section{Entanglement entropy \& massive flavors}
\label{massiveEE}

Moving one step forward from the setup we examined in the previous section, we will consider the strip in the Klebanov-Strassler with
massive dynamical flavors and calculate its EE.  The extra scale in the system, represented by the position that the source for the flavor branes
gets activated, will introduce a second peak in the function $L(\tau_0)$ that will correspond to a another first order phase transition for the EE.
The interplay/competition between the scales of the glueballs and the quarks will lead us to a critical point where one of the two phase transitions will become of the second order.
At the point where the first order phase transition degenerates to a second it is possible to calculate the critical exponents and determine the
universality class that the phase transition belongs. Even if we perform the calculation in the Heaviside approximation \eqref{step},
the critical point we observe is genuine and not an artifact of the approximation.

When dynamical flavors are present in the geometry, there is the notion of a screening length for the strip (in full analogy with the screening length in the Wilson loop computation,
e.g. \cite{Bigazzi:2008gd},\cite{Bigazzi:2008zt}). This is the length at which the connected surface would break by producing a pair of two cigars that end on a
some of the smeared flavor branes.
The value for that length would be determined by the condition that the EE for the connected surface equals the EE of two cigars ending on a flavor brane.
Due to the smearing of the flavors the decay of the EE into such a channel is $1/N_c$ suppressed, and not $N_f/ N_c$ as it would be for the case of the parallel localized flavor.

\begin{figure}[h] 
   \centering
   \includegraphics[width=7.5cm]{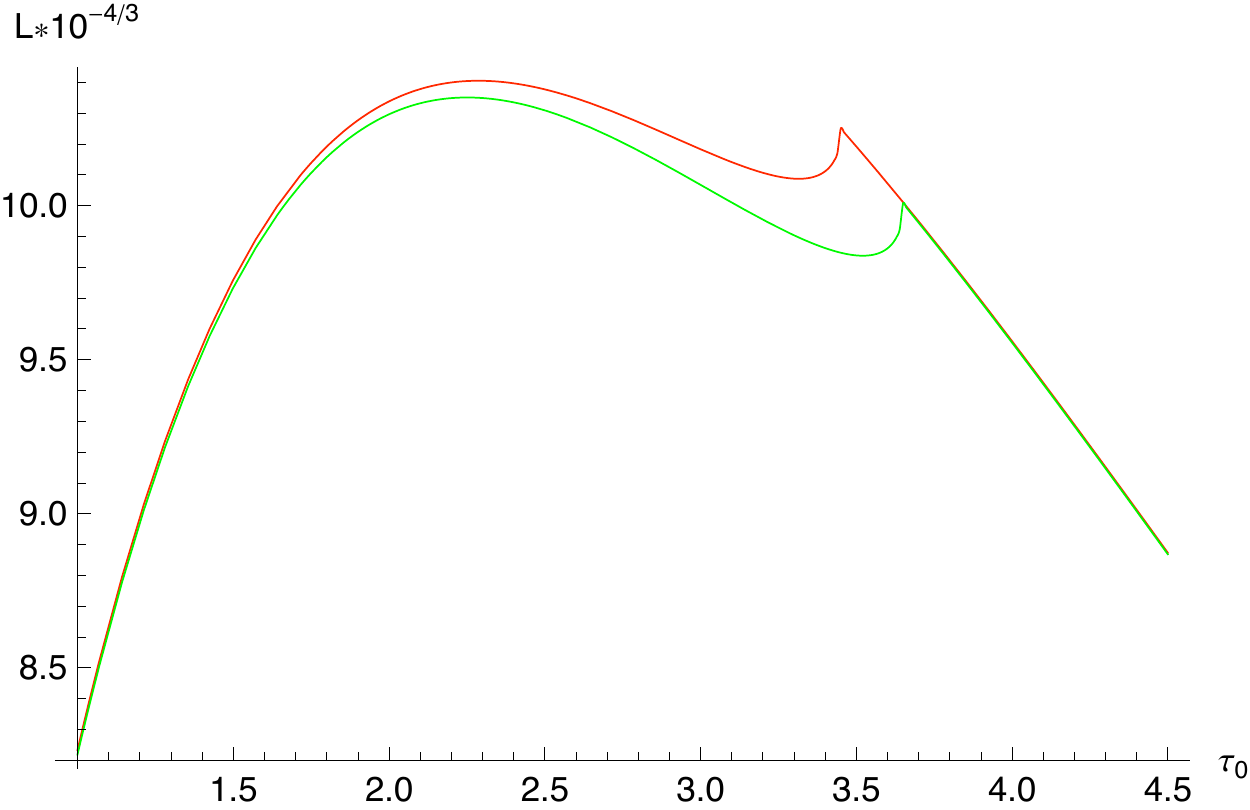}
   \includegraphics[width=7.5cm]{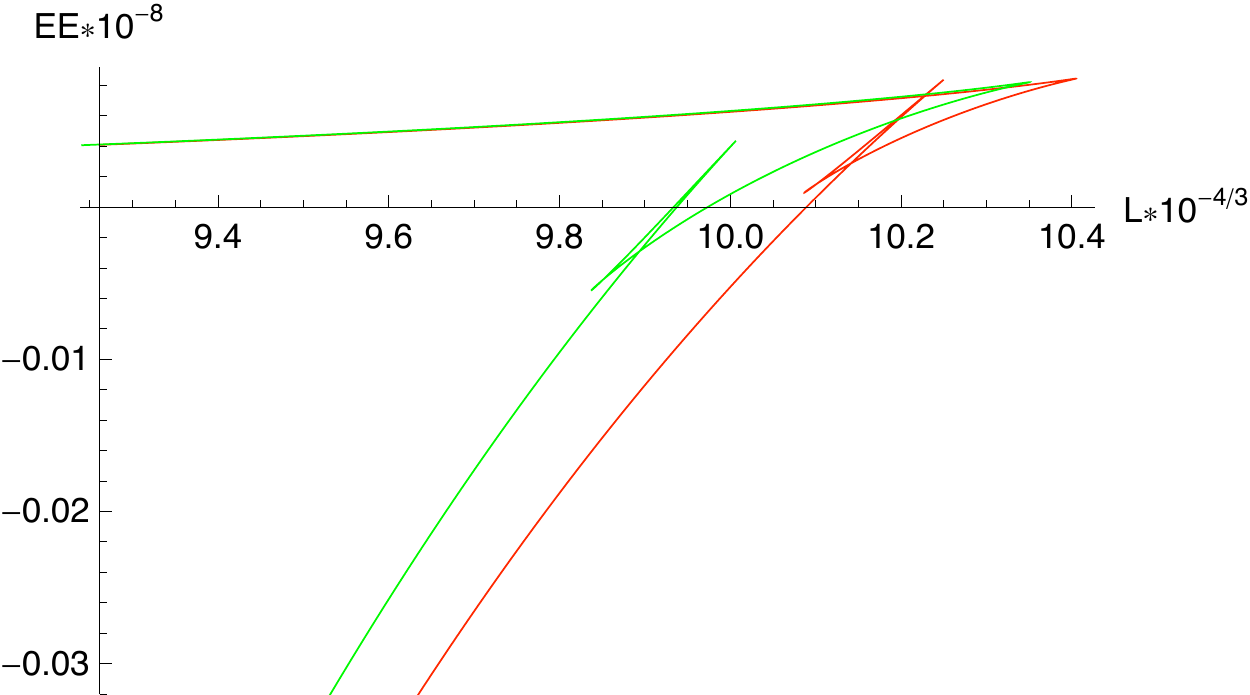}
   \caption{On the left part of the figure it is the plot of the length of the strip as a function of $\tau_0$ for two different radii of the spherical cavity.
    The red (green) curve is for $\mu=3.45$ ($\mu=3.65$).
    In the right part of the figure it is the plot of the EE as a function of $L$. In order for the physical branch
    of the EE to contain two phase transitions
    (the lower branch of the second phase transition that corresponds to the disconnected solution is for $EE=0$)
    the second peak should me noticeably lower than the first one, green vs. red
   (see also the analysis with a variety of examples in the appendix B.4 of \cite{Kol:2014nqa}). Notice that as we decrease
    $\mu$ the spike related to the quark mass approaches and finally overtakes the one related to the glueball scale.
   The calculation is for $N_f=1$.}
   \label{fig:4}
\end{figure}

Since in the previous section we thoroughly examined the dependence of the EE to the number of flavors in this section we will focus on the effect of the quark mass.
For this reason we fix the number of flavors to the value  $N_f=1$ and plot the function $L(\tau_0, \tau_q \equiv \mu)$
(and the EE) for different values of the quark mass.
When the two scales, namely the glueball and the quark scale, are noticeably far from each other (green vs. red plot in figure \ref{fig:4}) the triangle
(visualizing the first order phase transition) that corresponds to the quark scale is in the 
physical branch of the EE. Indeed in  figure \ref{fig:4} the red triangle is
in the unphysical region (for ${\rm EE}<0$), while the green triangle in the physical. 

We should mention that by comparing the plots of the EE with respect to the length of the stripe,
for different number of flavors, we can see that the value of $\mu$ that corresponds to the critical value for the length increases when you increase the number of
flavors.

\begin{figure}[h] 
   \centering
   \includegraphics[width=7.5cm]{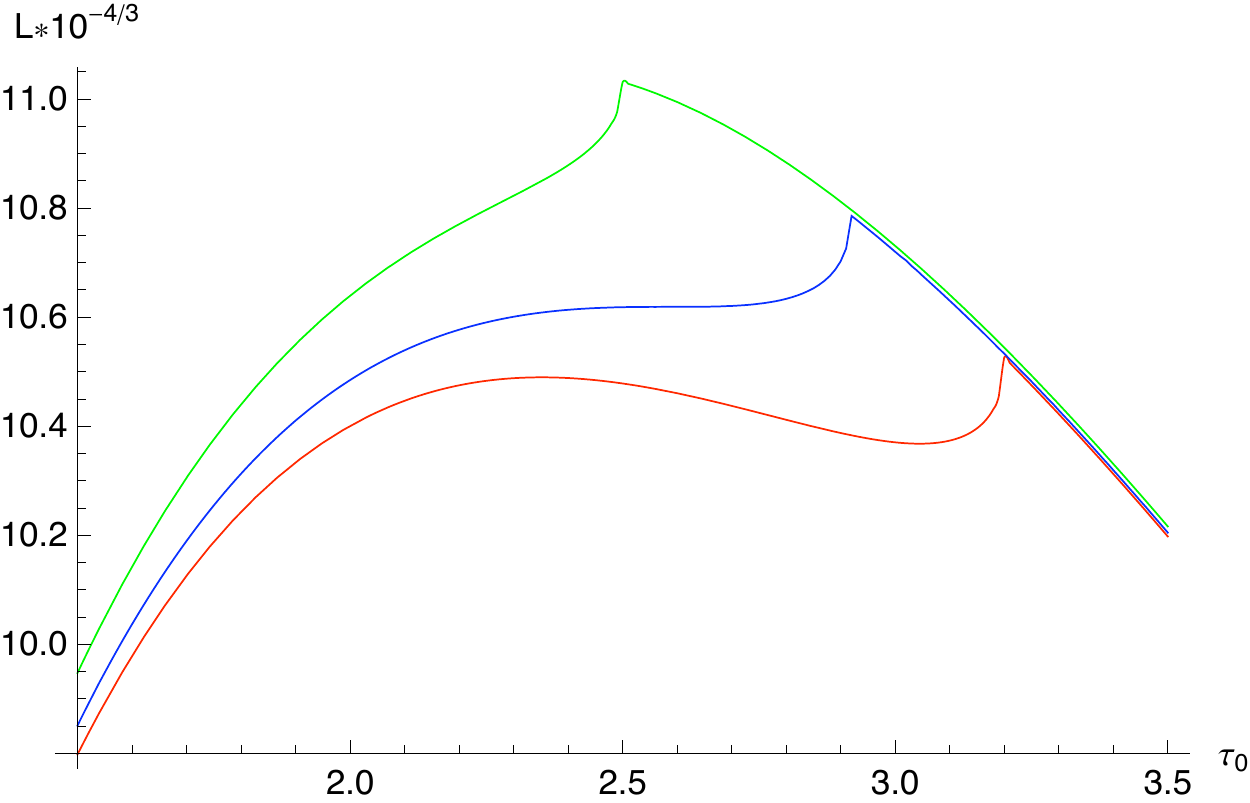}
   \includegraphics[width=7.5cm]{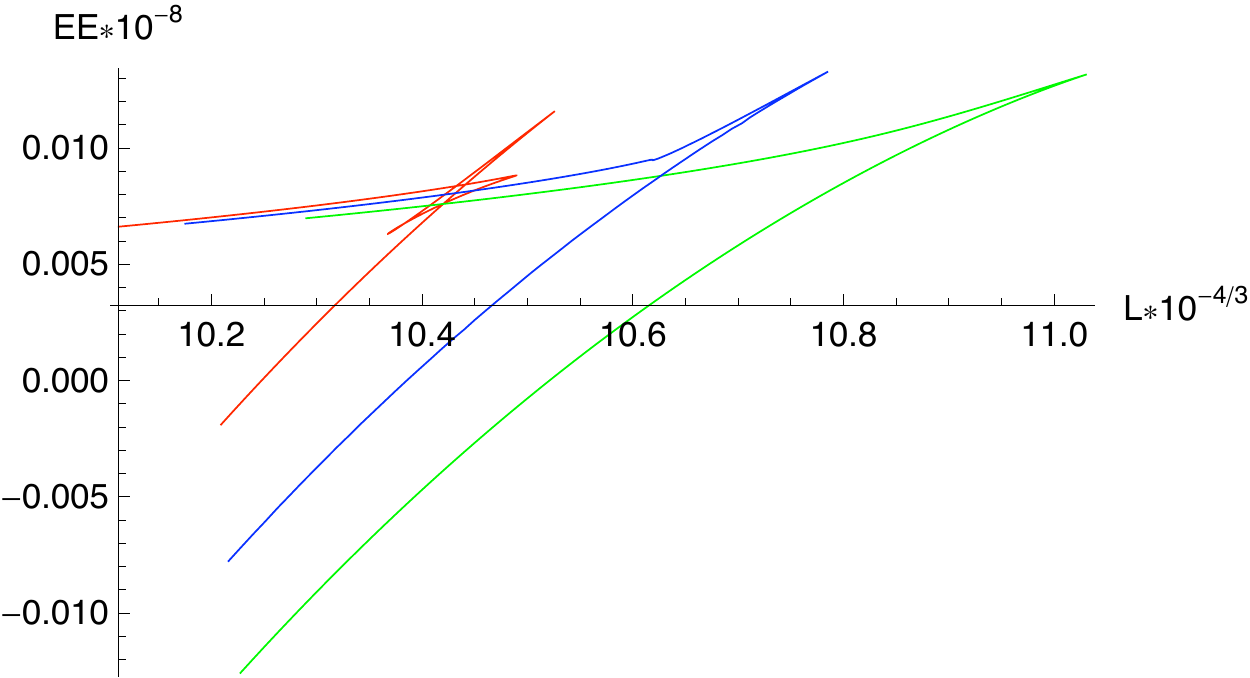}
   \caption{On the left part of the figure it is the plot of the length of the strip for three different radii of the spherical cavity.
    In the blue curve we have used the critical value of $\mu$, while the red (green) curve is for values of $\mu$ above (below) this critical value. The values we have used are
    $\mu=3.2$ \& $\mu=2.5$. In the right part of the figure it is the plot of the EE as a function of $L$. As can be seen from this plot the second triangle disappears exactly at the
    critical value of $\mu$ and if we keep on decreasing the critical value we are left with just one phase transition. The calculation presented in this plot is for $N_f=1$.}
   \label{fig:5}
\end{figure}

Keep decreasing the distance between the two energy scales we come to a point where one of the two peaks (the one that corresponds to the glueballs) is dissolved and
we are left with just one peak (see the left part of figure \ref{fig:5}). This peak corresponds to the energy scale set by the mass of the quarks.
The critical value of $\mu$ where the first peak disappears is a transition point between a first and a second order phase transition.
The calculation of the corresponding critical exponents will put us in the position to make a conjecture about the existence of a universality class for the transitions of the
EE\footnote{The calculation of the critical exponents is performed using the Heaviside approximation \eqref{step} for the flavor distribution function. Because of this,
as can be seen from the left plot of figure \ref{fig:4}, the peak that corresponds to the quark scale is more acute than the peak of the glueball scale.  Since the
peak that disappears, and for which we calculate the critical exponents, is the smooth one, the use of the heaviside approximation has no effect on the values of the
exponents and the determination of the universality class.}.

\begin{figure}[h] 
   \centering
   \includegraphics[width=7.5cm]{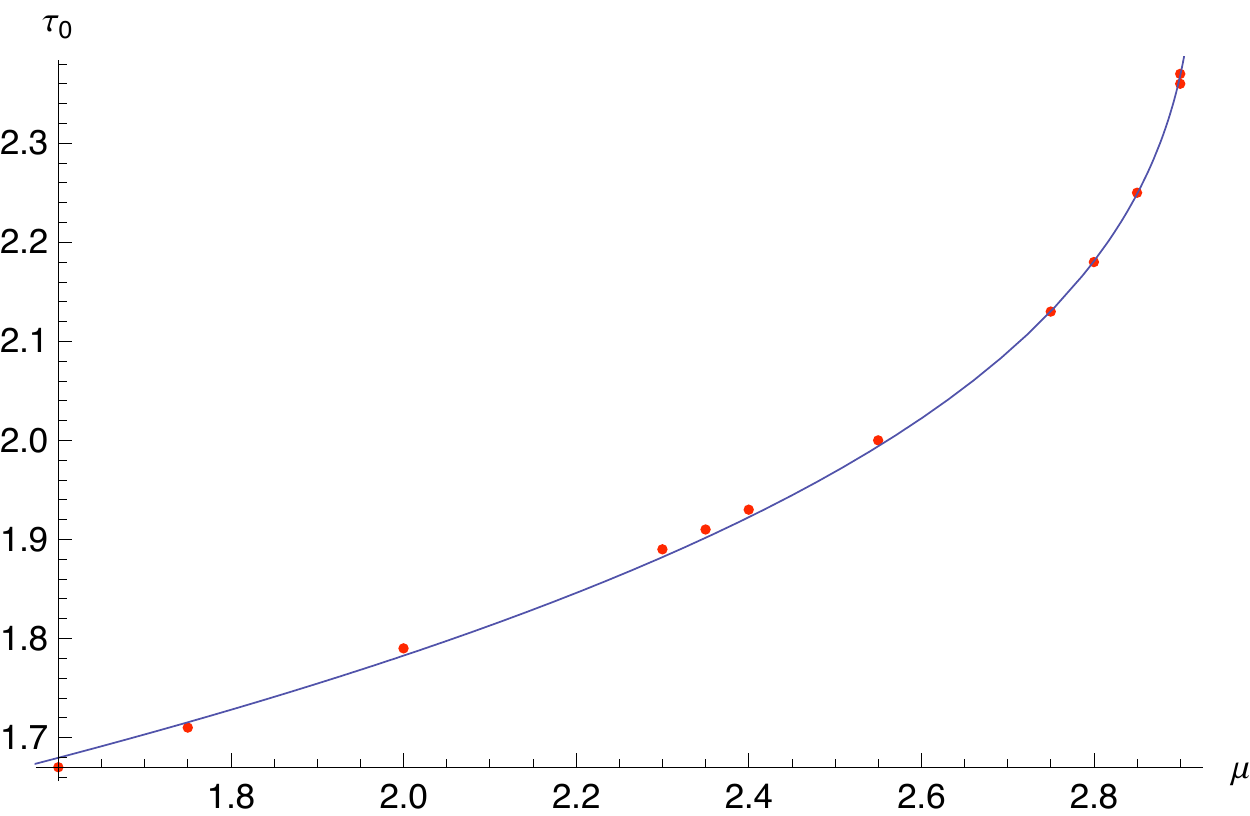}
   \includegraphics[width=7.5cm]{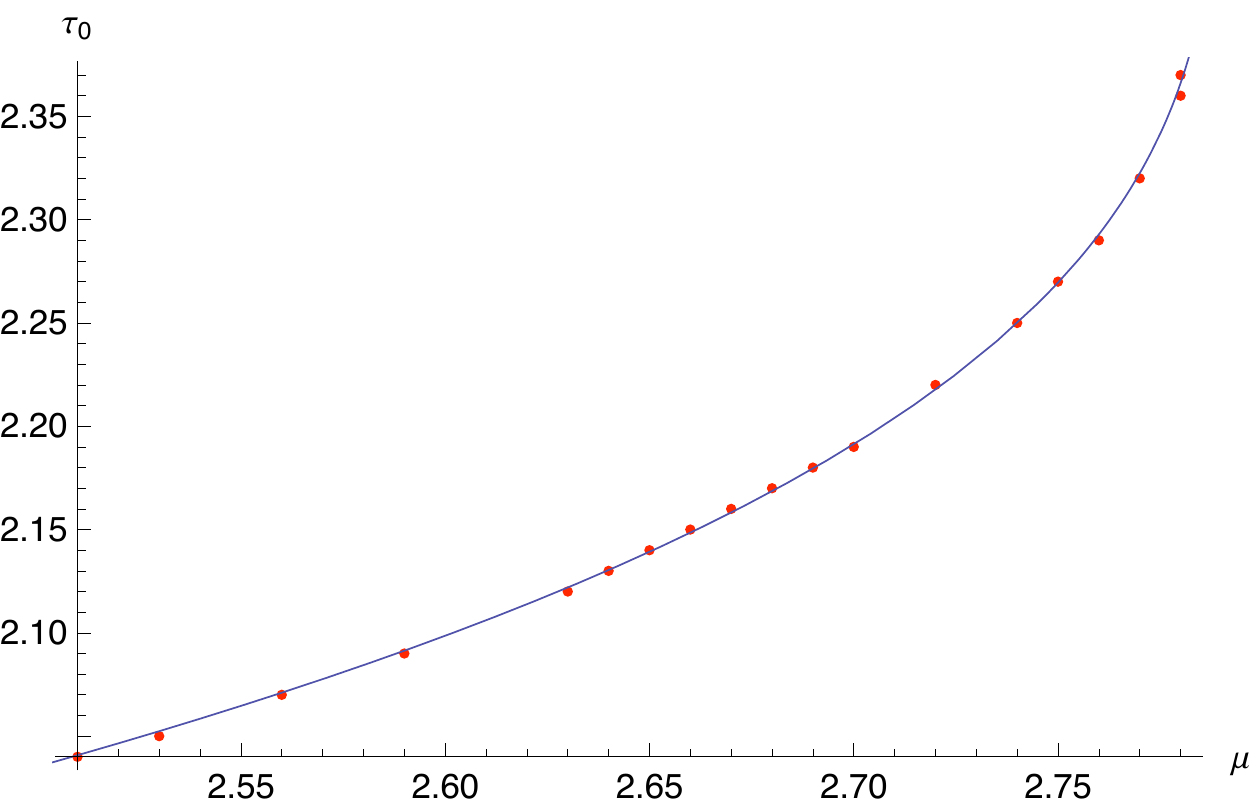}
   \caption{In the two plots of that figure we represent (with the solid blue curve) a non-linear fit to the data of the numerical computations for the minimal radial position in the bulk
   as a function of the quark mass. On the left (right) plot the data are for $N_f=1$ ($N_f=1/2$). In both cases the value of the critical exponent is universal, independent
   of the value of $N_f$ and different from the classical mean field theory value, namely $\beta= 1/3$.}
   \label{fig:6}
\end{figure}

In order to study and characterize the phase transition we will focus our analysis on the behavior of the function $L(\tau_0, \mu)$.  Plotting this function
we are looking for the critical value of $\mu$  where not only the first but also the second derivative of the function $L(\tau_0, \mu)$ with respect to $\tau_0$ vanishes.
Near the critical point the function  $L(\tau_0, \mu)$ will be well approximated by the following expression
\begin{equation} \label{Lapprox}
L \, - \, L_c\, \approx \, \left(\tau_0 \, - \, \tau_c \right)^3 \, .
\end{equation}
Performing the calculation after fixing the number of flavors to $N_f =1$ we get the following values for the critical length, $\mu$ \& $\tau$
\begin{equation}
L_c\, \approx \, 10.62 \, \quad \tau_c \, \approx \, 2.59  \quad \& \quad  \mu_c \, \approx \, 2.92 \, .
\end{equation}
From the behavior of the critical curve it is possible to extract the value for the critical exponents (for their definition see \cite{LeBellac:1991cq}).
Since along the critical value $\mu=\mu_c$ the behavior of the critical curve is $L \, - \, L_c\, \approx \, \left(\tau_0 \, - \, \tau_c \right)^{\delta}$,
the value for this critical exponent is $\delta=3$. In order to specify the critical exponent that is called $\beta$, we need to fit the numerical data to the following curve
\begin{equation} \label{tauapprox}
\tau_0 \, - \, \tau_c\, \approx \, \tau_c \left( 1 \, - \, {\mu \over \mu_c} \right)^{\beta} \, ,
\end{equation}
for $\mu<\mu_c$ at $L=L_c$. As can be seen from figure \ref{fig:6}, a non-linear fit (the solid blue curve) to the numerical data of the parametric plot of
the minimal radial position in the bulk  as a function of the quark mass, provides the critical value $\beta=1/3$ with a high accuracy. In order to
explore the possible dependence of this critical exponent to the number of flavors $N_f$, we performed the same computation for three different values of $N_f$,
namely $1/2$, $1$ \& $3/2$\footnote{In figure \ref{fig:6} we present the numerical fit for $N_f=1$ ($N_f=1/2$) corresponding to the left (right) plot.}.
Our findings to high accuracy indicate that the value  $\beta=1/3$ is universal and independent of the number of flavors.
Although the critical value for $\delta$ is the classical one (in agreement with the classical mean field theory value) the critical value for $\beta$ is not.
So far in the literature there are examples of calculations of critical exponents (in a variety of gravity backgrounds) that either
agree (see e.g \cite{Bigazzi:2008qq})
or disagree (see e.g. \cite{ Brandhuber:1999jr, Filev:2014mwa})
with the expectations that originate from the classical mean field theory value.
Thus, overall the behaviour of the strip width around the critical point takes the following form
\begin{equation} \label{L-tau-mu}
\frac{L}{L_c} \, - \, 1 \, = \, \left(\frac{\tau_0}{\tau_c}\, - \, 1 \right)^3 \, - \, \left(\frac{\mu}{\mu_c} \, - \, 1 \right) \, .
\end{equation}
Notice the difference between \eqref{L-tau-mu} and equation (5.1) of \cite{Bigazzi:2008qq}. In the latter the authors studied the heavy quark antiquark potential
(expectation value of two Wilson lines) in the massive KS background. They found the phase transition in the universality class of the Van der Waals system
with mean field theory critical exponents. They argued that this is  related to the theory of singularity of families
of functions, or “Catastrophe theory”, and that this universality reflects the fact that they were
analyzing a classical object: a macroscopic string. In our case the basic object is still classical, a minimal surface, but the system belongs in a different universality class,
probably because we are dealing with a higher dimensional surface.
A final comment is in order. Let us stress that the first order phase transition of \cite{Bigazzi:2008qq,Bigazzi:2008gd}
which is related to the existence of dynamical quarks is present only for values of $\mu$ smaller than a critical value. When $\mu$
is greater than this critical value their phase transition ceases to exist.
In our case the quark related phase transition is present for all values of $\mu$ becoming dominant for $\mu<\mu_{cr}$.
However, one should always keep in mind that we are considering different observables. We focus on the EE while the aforementioned studies focus
on the heavy quark potential.

Two of the critical exponents, through the use of the following scaling relations
\begin{equation} \label{scaling}
\alpha+2\beta+\gamma=2\ , \qquad \gamma=\beta(\delta-1)\ , \qquad 2-\alpha=\nu d\ , \qquad \gamma=\nu(2-\eta)\, ,
\end{equation}
are sufficient to completely determine the universality class of the system, in case we know its dimension $d$.
This dimension is related to the computation of the correlation function and since this calculation is beyond the scope of the paper, we will restrict on the
critical exponents $\alpha$ \& $\gamma$. Using \eqref{scaling} and the already known values for $\delta$ \& $\beta$ we have
\begin{equation} \label{criticalvalues}
\alpha \, = \, \gamma \, = \, \frac{2}{3} \, , \qquad \beta \, = \, \frac{1}{3} \, \qquad \& \qquad \delta \, = \, 3 \, .
\end{equation}

Finally, let us comment on the behaviour of the EE as a function of the width of the strip $L$ around the critical point.
As discussed above at $\mu=\mu_c$ the first order phase transition becomes a second order one with $\tau_0$ playing the role of the order parameter.
If we wanted to make a comparison with a thermodynamic system, the EE corresponds to the Gibbs free energy, the length of the strip corresponds to
the pressure, the order parameter $\tau_0$ corresponds to the density while the quark mass $\mu$ corresponds to the temperature.
As in the the quark antiquark potential one can use \eqref{length} and \eqref{EE_difference} to prove the following relations
\begin{equation} \label{conditions}
\frac{dS}{dL} \, = \, \frac{\sqrt{H(\tau_0)}}{2}>0 \qquad \& \qquad
\frac{d^2S}{dL^2} \, = \,  \frac{H'(\tau_0)}{4\sqrt{H(\tau_0)} \, L'(\tau_0)} \, .
\end{equation}
Near the critical point the function  $L(\tau_0)$ is well approximated by \eqref{Lapprox}. Substituting the derivative of \eqref{Lapprox} in the
second relation of \eqref{conditions} and replacing $\tau_0-\tau_c$ in terms of $L-L_c$ we can integrate to get the following expression
\begin{equation} \label{S-L}
S \, - \, S_c \, = \,  \frac{\sqrt{H(\tau_0)}}{2} \,  \left(L \, - \, L_c\right) \, \left(1 \, - \, D |L-L_c|^{1/3} \right) \, ,
\end{equation}
where $D$ is a constant that depends on $L, H$ and their derivatives at the critical point $\tau_c$.


\section{Conclusions \& outlook}
\label{conclusions}

In this paper we studied the EE of the confining KS background, when it is coupled to a large number of dynamical flavors (massive \& massless) in the
Veneziano limit. Dividing one of the spatial directions into a line segment of length $L$ and its complement the EE between the two subspaces admits a first
order phase transition, when we vary the width of the strip $L$.

In section \ref{masslessEE} we analyze the behavior of both the length of the strip and the EE as we vary the number of massless flavors.
Depending on the way we fix the integration constant of the warp factor, that is related to the glueball scale, we find two different quantitative behaviors.
Requiring the warp factor to vanish at some finite UV cut-off (in analogy with \cite{Klebanov:2000hb}), the glueball mass increases with the number of flavors.
The opposite behavior for the glueball mass occurs, when we impose on the warp factor to be the same at the origin for any value of $N_f$.

In section \ref{massiveEE} the mass of the flavor branes introduce another scale in the background. In this case the plot of the
length of the strip presents a second peak at the position that the flavor branes gets activated and the EE presents a second phase transition.
The competition between the glueball and the quark scales will lead to a critical point where the phase transition which was originally
associated with the glueball scale will become a second order one.
The critical exponents we calculate are independent of the number of flavors and different from the mean filed theory expectations.
This is in contradistinction with similar calculations of the quark-antiquark potential in the same background in \cite{Bigazzi:2008qq}.
On general grounds, since we consider the effect of the fundamental loops in the Feynman diagrams of the gauge theory, 
such a deviation from the mean field theory expectations for the critical exponents should not be unexpected.
Even in the quenched approximation quantum effects are sufficiently strong for that, see e.g. \cite{Filev:2014mwa}.\\
Finally, let us briefly comment on the physical picture behind our findings. When the mass of the quarks is much larger that the lightest glueball scale,
i.e. when $\mu>>\mu_{cr}$, the first order transition related to the existence of the quarks is happening closer to the UV cut-off
and affects little the IR physics, that is the QCD scale $\Lambda_{QCD}$ or the lightest glueball mass.
But as we decrease the quark mass the effect of
the quarks becomes significant since we can no longer integrate them out. Indeed, as can be seen from Figures \ref{fig:4} and \ref{fig:5}
when we decrease $\mu$ the critical length
$L_{cr}$ increases and as a result the QCD scale $\Lambda_{QCD}\sim 1/L_{cr}$ gets smaller. This qualitative behaviour can be understood if we consider
the two extreme limits. When $\mu \gg \mu_{cr}$ the quarks are very heavy and they decouple from the dynamics of the theory. Effectively this corresponds
to setting $N_f=0$. On the other hand, when $\mu \sim \mu_{cr}$ the presence of a large number of quarks increases the $\beta$-function 
which results in a smaller QCD scale.

We will close this section by describing a couple of ideas for future studies.
In \cite{Kol:2014nqa} they presented a detailed study of the EE for several confining theories,
with and without flavor. Some of the gravity models they examine, present multiple phase transitions in the EE, due to the presence of various scales
(e.g. quarks with different masses). One could try to study the behavior of the gravity system when the scales (masses) become comparable, by calculating the critical exponents.
It would be interesting to check whether those systems belong to the same universality class with the one we describe in this paper.
Since both peaks of the length of the strip will correspond to quark masses, another calculation would be to examine the possible dependence of the
critical exponents on the number of flavors of the different spherical cavities.

In \cite{Filev:2014nza} the construction of a gravity background
with backreacted flavor branes of different masses in the $AdS_5 \times S^5$ was presented. The flavor branes produce multiple spherical cavities with radii
that correspond to the quark masses. Generalizing that construction to the KS background will imply a confining model with multiple scales and multiple phase transitions.
It would be interesting to study the interplay of the different scales in the EE and the values of the critical exponents.


\section*{Acknowledgements}

We would like to thank V. Filev and especially C. N\'u\~nez for their useful comments and suggestions.
D.~Z.~is funded by the FCT fellowship SFRH/BPD/62888/2009. The work of G.G. was partially supported
by the General Secretariat for Research and Technology of Greece and from
the European Regional Development Fund MIS-448332-ORASY (NSRF 2007-13 ACTION,
KRIPIS).






\begin{thebibliography}{99}

\bibitem{Maldacena:1997re}
  J.~M.~Maldacena,
  ``The Large N limit of superconformal field theories and supergravity,''
  Int.\ J.\ Theor.\ Phys.\  {\bf 38}, 1113 (1999)
  [Adv.\ Theor.\ Math.\ Phys.\  {\bf 2}, 231 (1998)]
  [hep-th/9711200].

\bibitem{Ramallo:2013bua}
  A.~V.~Ramallo,
  ``Introduction to the AdS/CFT correspondence,''
  Springer Proc.\ Phys.\  {\bf 161}, 411 (2015)
  [arXiv:1310.4319 [hep-th]].


\bibitem{CasalderreySolana:2011us} 
  J.~Casalderrey-Solana, H.~Liu, D.~Mateos, K.~Rajagopal and U.~A.~Wiedemann,
  ``Gauge/String Duality, Hot QCD and Heavy Ion Collisions,''
  arXiv:1101.0618 [hep-th].


\bibitem{Edelstein:2009iv} 
  J.~D.~Edelstein, J.~P.~Shock and D.~Zoakos,
  ``The AdS/CFT Correspondence and Non-perturbative QCD,''
  AIP Conf.\ Proc.\  {\bf 1116}, 265 (2009)
  [arXiv:0901.2534 [hep-ph]].


\bibitem{Beisert:2010jr}
  N.~Beisert, C.~Ahn, L.~F.~Alday, Z.~Bajnok, J.~M.~Drummond, L.~Freyhult, N.~Gromov and R.~A.~Janik {\it et al.},
  ``Review of AdS/CFT Integrability: An Overview,''
  Lett.\ Math.\ Phys.\  {\bf 99}, 3 (2012)
  [arXiv:1012.3982 [hep-th]].

\bibitem{3-point}
  K.~Zarembo,
  ``Holographic three-point functions of semiclassical states,''
  JHEP {\bf 1009}, 030 (2010)
  [arXiv:1008.1059 [hep-th]].
\\
  M.~S.~Costa, R.~Monteiro, J.~E.~Santos and D.~Zoakos,
  ``On three-point correlation functions in the gauge/gravity duality,''
  JHEP {\bf 1011}, 141 (2010)
  [arXiv:1008.1070 [hep-th]].
\\
  G.~Georgiou,
  ``Two and three-point correlators of operators dual to folded string solutions at strong coupling,''
  JHEP {\bf 1102}, 046 (2011)
  [arXiv:1011.5181 [hep-th]].
\\
  G.~Georgiou,
  ``SL(2) sector: weak/strong coupling agreement of three-point correlators,''
  JHEP {\bf 1109}, 132 (2011)
  [arXiv:1107.1850 [hep-th]].
\\
  G.~Georgiou, V.~Gili, A.~Grossardt and J.~Plefka,
  ``Three-point functions in planar N=4 super Yang-Mills Theory for scalar operators up to length five at the one-loop order,''
  JHEP {\bf 1204}, 038 (2012)
  [arXiv:1201.0992 [hep-th]].
\\
  Y.~Jiang, I.~Kostov, F.~Loebbert and D.~Serban,
  ``Fixing the Quantum Three-Point Function,''
  JHEP {\bf 1404}, 019 (2014)
  [arXiv:1401.0384 [hep-th]].


\bibitem{Klebanov:2000hb}
  I.~R.~Klebanov and M.~J.~Strassler,
  ``Supergravity and a confining gauge theory: Duality cascades and chi SB resolution of naked singularities,''
  JHEP {\bf 0008}, 052 (2000)
  [hep-th/0007191].

\bibitem{Benini:2007gx}
  F.~Benini, F.~Canoura, S.~Cremonesi, C.~Nunez and A.~V.~Ramallo,
  ``Backreacting flavors in the Klebanov-Strassler background,''
  JHEP {\bf 0709}, 109 (2007)
  [arXiv:0706.1238 [hep-th]].


\bibitem{Bigazzi:2008qq}
  F.~Bigazzi, A.~L.~Cotrone, A.~Paredes and A.~V.~Ramallo,
  ``The Klebanov-Strassler model with massive dynamical flavors,''
  JHEP {\bf 0903}, 153 (2009)
  [arXiv:0812.3399 [hep-th]].




\bibitem{Nunez:2010sf}
  C.~Nunez, A.~Paredes and A.~V.~Ramallo,
  ``Unquenched Flavor in the Gauge/Gravity Correspondence,''
  Adv.\ High Energy Phys.\  {\bf 2010}, 196714 (2010)
  [arXiv:1002.1088 [hep-th]].


\bibitem{smearing}
   A.~V.~Ramallo, J.~P.~Shock and D.~Zoakos,
  ``Holographic flavor in N=4 gauge theories in 3d from wrapped branes,''
  JHEP {\bf 0902}, 001 (2009)
  [arXiv:0812.1975 [hep-th]].
  \\
   D.~Arean, E.~Conde, A.~V.~Ramallo and D.~Zoakos,
  ``Holographic duals of SQCD models in low dimensions,''
  JHEP {\bf 1006}, 095 (2010).
   [arXiv:1004.4212 [hep-th]].
   \\
    M.~Ammon, V.~G.~Filev, J.~Tarrio and D.~Zoakos,
  ``D3/D7 Quark-Gluon Plasma with Magnetically Induced Anisotropy,''
  JHEP {\bf 1209}, 039 (2012)
  [arXiv:1207.1047 [hep-th]].
  \\
   N.~Jokela, J.~Mas, A.~V.~Ramallo and D.~Zoakos,
  ``Thermodynamics of the brane in Chern-Simons matter theories with flavor,''
  JHEP {\bf 1302}, 144 (2013)
  [arXiv:1211.0630 [hep-th]].
 \\
  G.~Itsios, V.~G.~Filev and D.~Zoakos,
  ``Backreacted flavor in non-commutative gauge theories,''
  JHEP {\bf 1306}, 092 (2013)
  [arXiv:1304.5211 [hep-th]].




\bibitem{Ryu:2006bv}
  S.~Ryu and T.~Takayanagi,
  ``Holographic derivation of entanglement entropy from AdS/CFT,''
  Phys.\ Rev.\ Lett.\  {\bf 96}, 181602 (2006)
  [hep-th/0603001].

\bibitem{Ryu:2006ef}
  S.~Ryu and T.~Takayanagi,
  ``Aspects of Holographic Entanglement Entropy,''
  JHEP {\bf 0608}, 045 (2006)
  [hep-th/0605073].


\bibitem{Klebanov:2007ws}
  I.~R.~Klebanov, D.~Kutasov and A.~Murugan,
  ``Entanglement as a probe of confinement,''
  Nucl.\ Phys.\ B {\bf 796}, 274 (2008)
  [arXiv:0709.2140 [hep-th]].



\bibitem{Kol:2014nqa}
  U.~Kol, C.~Nunez, D.~Schofield, J.~Sonnenschein and M.~Warschawski,
  ``Confinement, Phase Transitions and non-Locality in the Entanglement Entropy,''
  JHEP {\bf 1406}, 005 (2014)
  [arXiv:1403.2721 [hep-th]].

\bibitem{Conde:2011rg}
  E.~Conde, J.~Gaillard and A.~V.~Ramallo,
  ``On the holographic dual of $N=1$ SQCD with massive flavors,''
  JHEP {\bf 1110}, 023 (2011)
  [Erratum-ibid.\  {\bf 1308}, 082 (2013)]
  [arXiv:1107.3803 [hep-th]].



\bibitem{Filev:2011mt}
  V.~G.~Filev and D.~Zoakos,
  ``Towards Unquenched Holographic Magnetic Catalysis,''
  JHEP {\bf 1108}, 022 (2011)
  [arXiv:1106.1330 [hep-th]].


\bibitem{Erdmenger:2011bw}
  J.~Erdmenger, V.~G.~Filev and D.~Zoakos,
  ``Magnetic Catalysis with Massive Dynamical Flavours,''
  JHEP {\bf 1208}, 004 (2012)
  [arXiv:1112.4807 [hep-th]].


\bibitem{Magana:2012kh}
  A.~Magana, J.~Mas, L.~Mazzanti and J.~Tarrio,
  ``Probes on D3-D7 Quark-Gluon Plasmas,''
  JHEP {\bf 1207}, 058 (2012)
  [arXiv:1205.6176 [hep-th]].



\bibitem{Bigazzi:2008zt}
  F.~Bigazzi, A.~L.~Cotrone and A.~Paredes,
  ``Klebanov-Witten theory with massive dynamical flavors,''
  JHEP {\bf 0809}, 048 (2008)
  [arXiv:0807.0298 [hep-th]].


\bibitem{Bigazzi:2009gu}
  F.~Bigazzi, A.~L.~Cotrone, A.~Paredes and A.~V.~Ramallo,
  ``Screening effects on meson masses from holography,''
  JHEP {\bf 0905}, 034 (2009)
  [arXiv:0903.4747 [hep-th]].


\bibitem{Filev:2014nza}
  V.~G.~Filev and D.~Zoakos,
  ``Multiple backreacted flavour branes,''
  JHEP {\bf 1412}, 186 (2014)
  [arXiv:1410.2879 [hep-th]].


\bibitem{Elander:2011mh}
  D.~Elander, J.~Gaillard, C.~Nunez and M.~Piai,
  ``Towards multi-scale dynamics on the baryonic branch of Klebanov-Strassler,''
  JHEP {\bf 1107}, 056 (2011)
  [arXiv:1104.3963 [hep-th]].



\bibitem{Kim:2013ysa}
  N.~Kim,
  ``Holographic entanglement entropy of confining gauge theories with flavor,''
  Phys.\ Lett.\ B {\bf 720}, 232 (2013).



\bibitem{HoyosBadajoz:2008fw}
  C.~Hoyos-Badajoz, C.~Nunez and I.~Papadimitriou,
  ``Comments on the String dual to N=1 SQCD,''
  Phys.\ Rev.\ D {\bf 78}, 086005 (2008)
  [arXiv:0807.3039 [hep-th]].

\bibitem{Bigazzi:2008gd}
  F.~Bigazzi, A.~L.~Cotrone, C.~Nunez and A.~Paredes,
  ``Heavy quark potential with dynamical flavors: A First order transition,''
  Phys.\ Rev.\ D {\bf 78}, 114012 (2008)
  [arXiv:0806.1741 [hep-th]].


\bibitem{LeBellac:1991cq}
  M.~Le Bellac,
  ``Quantum and statistical field theory,''
  Oxford, UK: Clarendon (1991) 592 p


\bibitem{Brandhuber:1999jr}
  A.~Brandhuber and K.~Sfetsos,
  ``Wilson loops from multicenter and rotating branes, mass gaps and phase structure in gauge theories,''
  Adv.\ Theor.\ Math.\ Phys.\  {\bf 3}, 851 (1999)
  [hep-th/9906201].


\bibitem{Filev:2014mwa}
  V.~G.~Filev,
  ``A Quantum Critical Point from Flavours on a Compact Space,''
  JHEP {\bf 1408}, 105 (2014)
  [arXiv:1406.5498 [hep-th]].




\end{thebibliography}
\end{document}